\documentstyle[12pt]{article}
\def\be{\begin{equation}}
\def\ee{\end{equation}}
\def\lb{\label}
\def\O{{\Omega}}
\def\R{\hbox{\bf R}}
\def\A{\hbox{\bf A}}
\def\F{\hbox{\bf F}}
\def\1{\hbox{\bf 1}}

\begin{document}

\title{Quantum Group Covariant Noncommutative Geometry.
\thanks{Dubna Preprint JINR E2-94-38, extended and revised version of
the report presented on the XXIIth International Conference
on Differential Geometric Methods in Physics, Mexico,
September 20-25, 1993.}}

\author{A.P. Isaev\thanks{e-mail address: isaevap@theor.jinrc.dubna.su}\\
\it Bogolubov Theoretical Laboratory, JINR, Dubna, \\
\it SU-101 000 Moscow, Russia }

\date{}

\maketitle

\begin{abstract}
The algebraic formulation of the quantum
group covariant noncommutative geometry in
the framework of the $R$-matrix approach
to the theory of quantum groups is given.
We consider structure groups taking values in the quantum groups
and introduce the notion of the noncommutative
connections and curvatures transformed
as comodules under the "local" coaction
of the structure group which is exterior extension of $GL_{q}(N)$.
These noncommutative connections and curvatures
generate $ GL_{q}(N)$-covariant quantum algebras.
For such algebras we find combinations of the generators
which are invariants under the coaction of the
"local"  quantum group and one can formally
consider these invariants as
the noncommutative images of the Lagrangians for the
topological Chern-Simons models, non-abelian gauge theories and
the Einstein gravity.
We present also an explicit realization of such
covariant quantum algebras via the
investigation the coset construction
$GL_{q}(N+1)/(GL_{q}(N)\otimes GL(1))$.

\end{abstract}

\newpage

\section{Introduction.}
\setcounter{equation}0

Noncommutative geometry \cite{Con} has started to play a
significant role in the mathematical physics for last few years.
One of the nontrivial examples of the noncommutative geometry
is given by the quantum groups \cite{FRT,D-M}.
After the paper \cite{W}, the differential geometric
aspects of the theory of quantum groups
have been intensively investigated recently
(see e.g. \cite{WJ}-\cite{IPy}).
Using these investigations various approaches
to formulate quantum group gauge theories have been developed
\cite{IP1}-\cite{AA}.

In this paper we prolong the researches
of a quantum group covariant noncommutative geometry proposed in
\cite{IP1,Is}. In the Sect.2 we
describe how it is possible to revise the
usual commutative geometry
(related to the geometry of the principal fibre bundle)
and introduce differentials covariant under the special
quantum group co-transformation interpreted as a local (structure)
transformation. Here
the special quantum group is an exterior extension of
$GL_{q}(N)$. Then we define corresponding geometrical objects
such as noncommutative
1-form connections and curvature 2-forms.
We show that these noncommutative geometrical objects generate
$GL_{q}(N)$-covariant quantum algebras.
In the Sect.3 we discuss the noncommutative
geometry related to the coset space $GL_{q}(N+1)/(GL_{q}(N)
\otimes GL(1))$. This geometry yields
the nontrivial explicit example of the algebraical constructions considered
in the Sect.2. Then, in the Sect.4, we
compose from the generators of the $GL_{q}(N)$-covariant
quantum algebras the set of $GL_{q}(N)$-local invariants,
which could be considered as the noncommutative images of the
well known gauge invariant Lagrangians
(e.g. discrete gauge theories and Einstein gravity).
Some of these invariants are nothing but
noncommutative analogs of the Chern characters. We
would like to stress, however,
that this analogy with the conventional Lagrangians
is rather formal and, strictly speaking, it may not lead to the
$q$-deformations of the corresponding field theories.

We use the notation and methods of the paper \cite{FRT}
in which $R$-matrix formulation of the quantum groups have been
elaborated. Some further development \cite{IPy2} of the $R$-matrix
notation considerably simplifying the calculations is also employed.
According to
the results obtained in \cite{AA} one can
reformulate our algebraical construction of the noncommutative geometry
for the case of the unitary structure groups $U_{q}(N)$.
Moreover we believe that using Brzezinski theorem \cite{B}
(and a generalization of it on the braided case \cite{IV})
about exterior Hopf algebras one can apply our construction
to the case of any quasitriangular Hopf algebra with
bicovariant first order differential calculus.
In the Conclusion we briefly
discuss this possibility and make some other remarks.

\section{$GL_{q}(N)$-covariant derivatives, noncommutative connections
and curvature.}
\setcounter{equation}0

Let us consider a
$Z_{2}$--graded finite dimensional Zamolodchikov algebra
(denoted by  $\O_{Z}$) generated by the operators
$\{ e^{i}, \; (de)^{j} \} $, $(i,j=1,2, \dots , N)$ with the following
commutation relations:
\begin{equation}
\hbox{\bf R} e e' = ce e' \; , \;\;
(\pm) c\hbox{\bf R} (de) e' = e (de)' \; , \;\;
\hbox{\bf R} (de) (de)' = - {1\over c} (de) (de)' ,
\lb{2.1}
\end{equation}
where $e=e_{1}$ is a $q$-vector in the first space,
$e'=e_{2}$ is a $q$-vector in the second space,
$\hbox{\bf R} = P_{12}R_{12}$ is a matrix which acts in the first and second
spaces simultaneously,
$P_{12}=
\delta^{i_{1}}_{j_{2}} \delta^{i_{2}}_{j_{1}}$ is the permutation matrix
and
\begin{equation}
\lb{2.1a}
\begin{array}{c}
R_{12}=R^{i_{1},i_{2}}_{j_{1},j_{2}}=
\delta^{i_{1}}_{j_{1}} \delta^{i_{2}}_{j_{2}}(1+(q-1)\delta^{i_{1}i_{2}}) +
(q-q^{-1})\delta^{i_{1}}_{j_{2}} \delta^{i_{2}}_{j_{1}}
\Theta_{i_{1}i_{2}} , \; \\   \\
\Theta_{ij} = \{ 1 \; {\rm if} \; i>j , \; 0 \; {\rm if} \; i \leq j \}
\end{array}
\end{equation}
is the $GL_{q}(N)$ $R$--matrix satisfying the Hecke relation
($\lambda = q-q^{-1}$).
\begin{equation}
\hbox{\bf R}^{2}=\lambda \hbox{\bf R} + \1.
\lb{2.2}
\end{equation}
Here $\1$ is $(N^{2} \times N^{2})$ unit matrix.
We imply the wedge product in the multiplication
of the differential forms in the formulas (\ref{2.1})
(we also omit $\wedge$ in all formulas below).
One can recognize in the
relations (\ref{2.1})
(for $(\pm)=+1$) the Wess-Zumino formulas of the covariant
differential calculus on  the bosonic ($c=q$)
and fermionic ($c=-1/q$) quantum hyperplanes \cite{WZ} where
$ e^{i} $ are the coordinates of the quantum hyperplane while
$(de)^{i}$ are the associated differentials.
The choice $(\pm)=-1$ corresponds to the case when $e^{i}$ are
bosonic $(c=-1/q)$ and fermionic $(c=q)$ veilbein 1-forms.
Note, that there is the second version of the algebra
(\ref{2.1}) obtaining by means of the replacement
$\R \rightarrow \R^{-1}, \; c \rightarrow c^{-1}$.
Below, we concentrate only on the consideration of the algebra
(\ref{2.1}) (the other type can be treated analogously).

It has been suggested in \cite{Man1,Sch,IP1} to consider the algebra
$\O_{Z}$ (\ref{2.1}) as a comodule with respect to the coaction
of the $Z_{2}$-graded quantum group $\O_{GL_{q}(N)}$
with the $GL_{q}(N)$-generators
$\{T^{i}_{j} \}$ and additional generators $\{ (dT)^{k}_{l}\}$
$ (i,j,k,l=1,2,....,N)$  which are
the basis of the differential 1-forms on the
quantum group $GL_{q}(N)$.
This coaction
$
\O_{Z}  \stackrel{g_{l}}{\longrightarrow} \O_{GL_{q}(N)} \otimes \O_{Z}
$
conserves the grading and can be written down as a homomorphism:
\begin{equation}
\lb{2.4a}
e^{i} \stackrel{g_{l}}{\longrightarrow} \
\widetilde{e}^{i} = T^{i}_{j} \otimes e^{j},
\end{equation}
\begin{equation}
\lb{2.4b}
(de)^{i} \stackrel{g_{l}}{\longrightarrow}
(\widetilde{de})^{i} = (dT)^{i}_{j} \otimes
e^{j} + T^{i}_{j} \otimes (de)^{j} .
\end{equation}
Here $\otimes$ denotes the graded tensor product:
$
a \otimes b = (-1)^{\hat{a}\hat{b}}(1 \otimes b)(a \otimes 1) \; ,
$
where $\hat{f} =deg(f)$ and $a \in \O^{(\hat{a})}_{GL_{q}(N)} \; , \;
b \in \O^{(\hat{b})}_{Z}$.
We recall that the algebra $\O_{Z}$ with the generators (\ref{2.1})
has the following expansion $\O_{Z} =
 \bigoplus\limits_{n=0}\Omega^{(n)}_{Z}$,
where $\Omega^{(n)}_{Z}$ denotes the subspace
of the differential n-forms and there exists a similar
expansion for the $Z_{2}$-graded quantum group
$\O_{GL_{q}(N)} =
 \bigoplus\limits_{n=0}\Omega^{(n)}_{GL_{q}(N)}$.
Substituting transformed algebra
$\{ \widetilde{e}^{i} , \; (\widetilde{de})^{i} \}$ into the commutation
relations (\ref{2.1}) we obtain the following equations
for the generators $\{ T_{j}^{i}, \; (dT)^{i}_{j} \} $
\begin{equation}
(\hbox{\bf R} - c)TT'(\hbox{\bf R} + c^{-1}) = 0 \; , \;\;
(\hbox{\bf R} (dT)T' - T(dT)' \hbox{\bf R}^{-1})(\hbox{\bf R} + c^{-1}) = 0
\; ,
\lb{2.6}
\end{equation}
\begin{equation}
(\hbox{\bf R} + c^{-1})(dT)(dT)'(\hbox{\bf R} + c^{-1}) = 0 \; , \;\;
(\hbox{\bf R} + c^{-1})
( (dT)T' \hbox{\bf R} - \hbox{\bf R}^{-1} T(dT)' ) =0 \; ,
\lb{2.7}
\end{equation}
where $T = T_{1} = T \otimes I$ while
$T' = T_{2} = I \otimes T$ and $I$ is a $(N \times N)$ unit matrix.
The relations
(\ref{2.6}), (\ref{2.7})
have to be fulfilled as for $c=q$ as for
$c = -q^{-1}$ therefore we deduce from them
the following $q$-commutation relations for the bicovariant differential
complex on $GL_{q}(N)$ (see \cite{Sch,Z,IPy}):
\begin{eqnarray}
\hbox{\bf R} T T' & = & T T' \hbox{\bf R} \; ,
\lb{2.10} \\
\hbox{\bf R} (dT) T' & = & T (dT)' \hbox{\bf R}^{-1} \; ,
\lb{2.11}
 \\
\hbox{\bf R} (dT) (dT)' & = & - (dT) (dT)' \hbox{\bf R}^{-1} \; .
\lb{2.12}
\end{eqnarray}
We stress that (\ref{2.12}) follows from (\ref{2.11}) if
the differential $d$ is nilpotent
$d^{2}=0$ and obeys the graded Leibnitz rule
$d(fg) = d(f)g + (-1)^{\hat{f}} f d(g)$.
It is interesting to note (see \cite{IP1})
that the algebra  $\O_{GL_{q}(N)}$
(\ref{2.10})-(\ref{2.12}) is the Hopf algebra.
The comultiplication
$\Delta$, the counit $\epsilon$   and the
antipode $S$ are defined by
\begin{equation}
\begin{array}{c}
\Delta (T) = T \otimes T \ , \;\;
\epsilon(T) = I \  , \;\;
{\cal S} (T) = T^{-1} \ , \cr
\Delta(dT) = dT \otimes T + T \otimes dT
\ , \;\; \epsilon(dT) = 0 \  ,\;\;
 {\cal S} (dT) = - T^{-1} dT T^{-1} ,
\end{array}
\lb{2.13}
\end{equation}
and satisfy to the all axioms of the Hopf algebra.
Thus, the algebra $\Omega_{GL_{q}(N)}$ yields the special
example of the general exterior Hopf algebras discussed
in \cite{B}. We stress that the example of $GL_{q}(N)$-exterior
Hopf algebra proposed in \cite{B} has slightly different
comultiplication comparing with the Hopf algebra
$\Omega_{GL_{q}(N)}$ (\ref{2.10})-(\ref{2.13}) independently
introduced in \cite{IPy,IP1}.
One can show that it is possible to extend
the action of the differential $d$ over the tensoring
and apply $d$ to the algebra $\O_{GL_{q}(N)} \otimes \O_{Z}$
in such a way that:
$
d(g \otimes \O_{Z}) = d(g) \otimes \O_{Z} +
(-1)^{k} g \otimes d(\O_{Z}),
$
where $g \in \O^{(k)}_{GL_{q}(N)}$ and $d^{2}=0$.

Now we would like to
interpret the formulas (\ref{2.4a}) and (\ref{2.4b})
as a local (structure) quantum group
transformation of the comodule $e^{i}$.
Here the matrix $T^{i}_{j}$
is understood as a noncommutative
analog of a local (structure) group element.
In view of this, it is natural to consider the
appearing of the additional
term $(dT)^{i}_{j} \otimes e^{j}$ in (\ref{2.4b})
as a noncovariance of the comodule
$(de)^{i}$ under the "gauge" rotation (\ref{2.4a})
(or as an indication that the differentials $(de)^{i}$
describe "nonparallel transporting" of the vector $e^{i}$).
To restore the covariance let us introduce a covariant
differential $\nabla$ in such a way that the transformations
(\ref{2.4a}), (\ref{2.4b}) are rewritten in the form
\begin{equation}
\lb{2.17a}
e^{i} \stackrel{g_{l}}{\longrightarrow} \
\widetilde{e}^{i} = T^{i}_{j} \otimes e^{j},
\end{equation}
\begin{equation}
\lb{2.17b}
(\nabla e)^{i} \stackrel{g_{l}}{\longrightarrow}
\widetilde{(\nabla e)}^{i}
= T^{i}_{j} \otimes (\nabla e)^{j} \; .
\end{equation}
In general $(\nabla e)^{i} \in \!\!\!\!\! / \O_{Z}$ and,
hence, the action of the operator $\nabla$ enlarges the algebra
$\O_{Z}$ up to some new algebra $\O_{\bar{Z}}$.
The operator $d$ can be induced (as a differential) onto
the whole algebra $\O_{\bar{Z}}$ and this algebra is
naturally decomposed as
$\O_{\bar{Z}} = \bigoplus_{n=0} \Omega^{(n)}_{\bar{Z}} ,$
where $\Omega^{(n)}_{\bar{Z}}$ is the subspace of $n$-forms.
Then we postulate that the elements
$(\nabla e)^{i} \in \Omega^{(1)}_{\bar{Z}}$
are expanded over the generators $\{ e^{i}, \; (de)^{j} \}$
of $\Omega_{Z}$ in the following way:
\begin{equation}
\lb{2.19}
(\nabla e)^{i} = (de)^{i} - A^{i}_{j} e^{j} ,
\end{equation}
It is clear that the coefficients $A^{i}_{j}$
belong to the subspace $\O_{\bar{Z}}^{(1)}$
and it is natural to consider them as noncommutative analogs
of the connection 1-forms. Under the transformations
(\ref{2.17a}) and (\ref{2.17b})
1-forms $A^{i}_{j}$ are transformed as:
\begin{equation}
\lb{2.20}
A^{i}_{k} \stackrel{g_{l}}{\longrightarrow}
\widetilde{A^{i}_{k}} =
T^{i}_{j} (T^{-1})^{l}_{k} \otimes A^{j}_{l}
+ dT^{i}_{j} (T^{-1})^{j}_{k} \otimes 1
\equiv
(T A T^{-1})^{i}_{k} + (dT T^{-1})^{i}_{k}
\; ,
\end{equation}
Here $\widetilde{A}^{i}_{j} \in \O_{GL_{q}(N)} \otimes \bar{Z} $.
In the last part of (\ref{2.20}) a short notation
is introduced to be used below.
The second action of the covariant derivative $\nabla$ on the
expression (\ref{2.19}) leads to the definition of the
curvature 2-forms $F^{i}_{j} \in \O^{(2)}_{\bar{Z}}$:
\begin{equation}
\lb{2.21}
\nabla(\nabla e) = - \left( d(A) - A^{2}\right) e = - Fe .
\end{equation}
The quantum co-transformation (\ref{2.20}) for the curvature
2-forms $F^{i}_{j}$ is represented as the adjoint coaction:
\begin{equation}
\lb{2.21a}
F^{i}_{j}{ \stackrel{g_{ad}}{\longrightarrow}}
\ \widetilde{F}^{i}_{j}
=  T^{i}_{k} (T^{-1})^{l}_{j}  \otimes F^{k}_{l}
\equiv T^{i}_{k}F^{k}_{l}(T^{-1})^{l}_{j} \; .
\end{equation}
The curvature tensor $F^{i}_{j}$ is a reducible adjoint representation
of $GL_{q}(N)$ and it is possible to decompose it into the
scalar $F^{0} = Tr_{q}(F)$ and the $q$-traceless tensor:
$$
\tilde{F}^{i}_{j} = F^{i}_{j} - \delta^{i}_{j} Tr_{q}(F)/ Tr_{q}(I).
$$
Here, we have introduced the $q$-deformed trace \cite{FRT,Z,IP1,IM}
for the case of the $GL_{q}(N)$-group
\begin{equation}
\lb{2.20a}
F^{0} = Tr_{q}(F) \equiv Tr(DF) \equiv \sum^{N}_{i=0} q^{-N-1+2i} F^{i}_{i}.
\end{equation}
Below we need the feature of invariance of $q$-trace:
\begin{equation}
\lb{inv}
Tr_{q}(TET^{-1})=Tr_{q}(E)
\end{equation}
where
$[T_{ij}, \; E_{kl} ] =0$ and
$T^{i}_{j} \in GL_{q}(N)$.
In particular, we have
\begin{equation}
\lb{4.3?}
Tr_{q2}(\hbox{\bf R} E \hbox{\bf R}^{-1})
= Tr_{q2}(\hbox{\bf R}^{-1} E \hbox{\bf R}) = Tr_{q}(E)
\end{equation}
Here $Tr_{q2}(.)$ denotes quantum trace over the second space.
We also use the relations:
\begin{equation}
\lb{2.a20}
Tr_{q}(\hbox{\bf R}^{\pm 1}) = q^{\pm N}, \; Tr_{q}(I) =
\frac{q^{N} - q^{-N}}{q - q^{-1}} \equiv [N]_{q}
\end{equation}

The next action of the covariant derivative on the formula
(\ref{2.21}) yields the Bianchi identities that are represented
in the classical form:
$$
d(F)=[A , \; F] .
$$

To complete the definition of the algebra $\O_{\bar{Z}}$
we have to deduce the commutation relations of the new
generators $\{ A^{i}_{j}, \; F^{i}_{j}, \dots \}$ and their
cross-commutation relations
with the generators $\{ e^{i}, \; (de)^{j} \}$.
First of all, let us note that the choice of the connection
in the pure gauge form (see (\ref{2.20}))
\begin{equation}
\lb{2.22}
A^{i}_{j} = dT^{i}_{k}(T^{-1})^{k}_{j} \otimes 1 \; ,
\end{equation}
leads to the conclusion that the generators $A^{i}_{j}$ could
satisfy the following $q$-deformed anticommutation relations:
\begin{equation}
\lb{2.23}
\hbox{\bf R} \A \hbox{\bf R} \A
+ \A \hbox{\bf R} \A \hbox{\bf R}^{-1} = 0 \; ,
\end{equation}
where $\A = A_{1} = A \otimes I$.
These relations
for the noncommutative 1-form connections (gauge fields)
have been postulated in \cite{IP1,CW}.
Note, however, that in the right hand side of Eq.(\ref{2.23})
one may add the arbitrary linear combination of the
curvature 2-forms $F = dA - A^{2}$ which is vanished on the
solution (\ref{2.22}). Thus, the general
covariant commutation relations for $A^{i}_{j}$ are
\begin{equation}
\lb{2.24}
\hbox{\bf R} \A \hbox{\bf R} \A + \A \hbox{\bf R} \A \hbox{\bf R}^{-1}
= a(\hbox{\bf R}) (\F\hbox{\bf R} + \hbox{\bf R}^{-1}\F)
+ \kappa(\hbox{\bf R}) F^{0} \equiv \Delta(F) \; ,
\end{equation}
where $\F = F_{1} = F \otimes I$, $a(\hbox{\bf R})
= a_{1} + a_{2}\hbox{\bf R}$ and for convenience
we choose the parameter $\kappa(\hbox{\bf R})$ in the form:
$\kappa(\hbox{\bf R}) = (\kappa_{1} + \kappa_{2}\hbox{\bf R})(\hbox{\bf R}
+ \hbox{\bf R}^{-1})$.

The special form of the right hand side of Eq.(\ref{2.24})
is dictated by the symmetry properties of the
$q$-anticommutator appeared in the left hand side of
this equation ($c = \pm q^{\pm 1}$):
$$
(\hbox{\bf R} -c)(\hbox{\bf R} \A \hbox{\bf R} \A
+ \A \hbox{\bf R} \A \hbox{\bf R}^{-1})
(\hbox{\bf R} + c^{-1}) = 0 \; .
$$
We stress that the anticommutation relations (\ref{2.24}) are
covariant under the transformations (\ref{2.20}) and (\ref{2.21a}).
Moreover one can extract from the relations (\ref{2.24}) subsets
of covariant relations using the methods proposed in \cite{IPy2}.
%%%%%%%%%%%%%%%%%%%%%%%%%%%%%%%%%%%%%%%%%%%%%%%%%%%%%%%%%%%%%%%%%%
Namely, applying $Tr_{q(2)}(...)$ and $Tr_{q(2)}(...\hbox{\bf R})$ to
(\ref{2.24})
and using (\ref{2.a20})
we obtain two sets of relations transformed as adjoint comodules:
\begin{equation}
\lb{2.a}
\begin{array}{c}
\lambda q^{N} A^{2} + \{ A^{0}, \; A \} =
[ a_{1}(q^{N} + q^{-N}) + a_{2}( [N]_{q} + \lambda q^{N}) ]F
+a_{2}F^{0} + \\ \\ +
[ \kappa_{1}(q^{N} + q^{-N}) + \kappa_{2}( 2[N]_{q} + \lambda q^{N}) ]
F^{0} \; ,
\end{array}
\end{equation}
\begin{equation}
\lb{2.b}
\begin{array}{c}
 q^{N} A^{2} + ( A * A ) =
[ a_{1}( [N]_{q} + \lambda q^{N})  +
 a_{2}q^{N}(q^{2} + q^{-2}) ]F + (a_{1} + \lambda a_{2})F^{0} + \\ \\
+ \left[ \kappa_{1}( 2[N]_{q} + \lambda q^{N})  +
 \kappa_{2}(q^{N}(q^{2} + q^{-2}) + \lambda [N]_{q}) \right] F^{0} \; ,
\end{array}
\end{equation}
where $(A * A) = Tr_{q(2)}(\hbox{\bf R}\A\hbox{\bf R}\A\hbox{\bf R}),
\; F^{0} = Tr_{q}(F), \;
A^{0} = Tr_{q}(A)$.
Then, applying $Tr_{q(1)}(...)$ to (\ref{2.a}) and (\ref{2.b})
we obtain two scalar relations $(q^{2} \neq -1)$
\begin{equation}
\lb{2.c}
Tr_{q}(A^{2})  =
\left[ (a_{1} + \kappa_{1}) q^{-N}[N]_{q} + (a_{2} + \kappa_{2})
\right] F^{0} \; ,
\end{equation}
\begin{equation}
\lb{2.d}
(A^{0})^{2}  =
[ (a_{1} + \kappa_{1}) q^{-N} +
 (a_{2} + \kappa_{2}) [N]_{q} ] F^{0} \; .
\end{equation}
We see that in the noncommutative case
Eqs.(\ref{2.c})-(\ref{2.d}) give additional relations
of 1-form connections $A$ and 2-form curvatures $F \equiv dA -A^{2}$.
%%%%%%%%%%%%%%%%%%%%%%%%%%%%%%%%%%%%%%%%%%%%%%%%%%%%%%%%%%%%%%

Arbitrary parameters $a_{i}, \; \kappa_{i}$ introduced in Eq.(\ref{2.24})
depend on the choice of the noncommutative geometry and
have to be fixed partially by the consistency conditions
(with respect to the two ways of ordering of any cubic monomial)
for the algebra $\O_{\bar{Z}}$.
It is amusing to note that the additional
nonzero term included into the right-hand side of (\ref{2.24})
looks similar to the quantum anomaly terms arising in the
(anti)commutators of fields (or currents)
in certain conventional quantum field theories.

In order to find commutation relations $A^{i}_{j}$ with the generators
$\{ e^{i}, \; (de)^{j} \}$ we postulate that the coordinates
of the comodule (\ref{2.19}) commute in the same way
as the components of 1-forms $(de)^{i}$ (see (\ref{2.1}))
\begin{equation}
\lb{2.25a}
\hbox{\bf R} (\nabla e)(\nabla e)' =
- {1\over c} (\nabla e) (\nabla e)'
\end{equation}
\begin{equation}
\lb{2.25b}
(\pm)(c-b)\hbox{\bf R} (\nabla e) e' =
 e (\nabla e)' \; .
\end{equation}
where $b$ is a constant which is fixed below.
Let us stress that Eqs.(\ref{2.25a}),(\ref{2.25b}) are not the general
covariant relations of that kind. For example one can add to (\ref{2.25a})
the terms of the type $(Fe)e'$. We however prefer to consider here
the simplest case of the relations (\ref{2.25a}),(\ref{2.25b}).
 From (\ref{2.1})
and (\ref{2.25b}) we deduce covariant commutation relations
of $A$ and $e$:
\begin{equation}
\lb{2.26}
(\pm) e \A' = \hbox{\bf R}\A\hbox{\bf R} e + b \hbox{\bf R} (\nabla e)
\end{equation}
Considering the consistency condition for the reordering
(in two different ways)
the monomial $e e' \A'' \equiv e_{1} e_{2} A_{3}$ we
obtain only two solutions for the parameter $b$:
\begin{equation}
\lb{2.26a}
A.) \;\; b = 0 \;\;\;\; B.) \;\; b = \lambda \; .
\end{equation}
Thus, we have two variants for the Eq.(\ref{2.26})
\begin{equation}
\lb{2.27a}
A.) \;\; (\pm)e \A' = \hbox{\bf R}\A\hbox{\bf R} e \; , \;\;
B.) \;\; (\pm)e \A' =
\hbox{\bf R}\A\hbox{\bf R}^{-1} e + \lambda \hbox{\bf R} (de) \; .
\end{equation}
Note, that in the paper \cite{IP1} we have considered only the
first case A.): $b=0$.
Taking into account (\ref{2.25a}) one can obtain the
corresponding commutation relations for $(de)$ and $A$
\begin{equation}
\lb{2.28}
(\pm)(de) \A' = - \hbox{\bf R}^{-1}\A\hbox{\bf R} (de) +
(b-\lambda) \A\hbox{\bf R} (\nabla e) +
\tilde{a}(\hbox{\bf R}) \Delta(F) e \; ,
\end{equation}
where
\begin{equation}
\lb{2.29a}
\tilde{a}(\hbox{\bf R}) = \frac{1+ \gamma(\hbox{\bf R} -c)}{1 + c^{2}}
\end{equation}
and $\gamma$ is a new arbitrary constant to be fixed below.
Type A.) and type B.)
commutation relations (\ref{2.26}), (\ref{2.28}) are
covariant under the gauge coactions (\ref{2.4a}), (\ref{2.4b}) and
(\ref{2.20}) and both cases lead to the same covariant
commutation relation
for $(\nabla e)$ and $A$:
\begin{equation}
\lb{2.29}
 (\pm)(\nabla e) \A' = - \hbox{\bf R}\A\hbox{\bf R} (\nabla e) +
(\tilde{a}(\hbox{\bf R}) - 1 ) \Delta(F) e \; ,
\end{equation}
Differentiating (\ref{2.26}) and, then,
using (\ref{2.28}) one can derive
\begin{equation}
\lb{2.30}
\begin{array}{c}
 e \F ' =  \hbox{\bf R}\F (\hbox{\bf R} -b) e  +
\tilde{\tilde{a}}(\hbox{\bf R}) \Delta(F) e = \cr \cr
 = (\hbox{\bf R} + \tilde{\tilde{a}}(\hbox{\bf R})
a(\hbox{\bf R}) ) \F \hbox{\bf R} e +
(\tilde{\tilde{a}}(\hbox{\bf R}) a(\hbox{\bf R}) \hbox{\bf R}^{-1}
-b\hbox{\bf R}) \F e +
\tilde{\tilde{a}}(\hbox{\bf R}) \kappa(\hbox{\bf R}) F^{0} e
\end{array}
\end{equation}
where we define
\begin{equation}
\lb{2.300}
\tilde{\tilde{a}}(\hbox{\bf R}) =
-(1+b\hbox{\bf R}) \tilde{a}(\hbox{\bf R}) + (b-\lambda)\hbox{\bf R} .
\end{equation}
Considering the reordering of the monomials $ee'\F''$
in two possible ways and comparing the results we obtain
for both types A.) $b=0$ and B.) $b= \lambda$
the restrictions
\begin{equation}
\lb{2.301}
1.) \;\; \tilde{\tilde{a}}(\hbox{\bf R}) a(\hbox{\bf R}) =0 \; ,
 \;\; \tilde{\tilde{a}}(\hbox{\bf R}) \kappa(\hbox{\bf R}) =0 ,
\end{equation}
which leads to the commutation relation:
\begin{equation}
\lb{2.30a}
e \F' = \hbox{\bf R} \F (\hbox{\bf R} - b) e.
\end{equation}
Note, that for the type A.) $(b = 0)$ we have an additional solution
$$
2.)
\tilde{\tilde{a}}(\hbox{\bf R}) a(\hbox{\bf R}) =-\lambda \; , \;\;
\tilde{\tilde{a}}(\hbox{\bf R})
\kappa(\hbox{\bf R}) = 0
$$
equivalent to the relation:
$e \F' = \hbox{\bf R}^{-1} \F \hbox{\bf R}^{-1} e$.
This relation, however, is consistent with the algebra
(\ref{2.24}), (\ref{2.26}) and (\ref{2.29})
only if some additional relations on the generators of
$\Omega_{\bar{Z}}$ will be fixed.
One can prove this by considering
two different ways of reordering of the monomials
$e \hbox{\bf R}' \A' \hbox{\bf R}' \A'$ where $\hbox{\bf R}' = P_{23}R_{23}$.

Taking into account the conditions (\ref{2.301}) we
obtain from the definitions (\ref{2.300}) and (\ref{2.29a})
the following solutions for the
parameters $a(\hbox{\bf R})$ and $\gamma$
\begin{equation}
\lb{2.30c}
\begin{array}{l}
1.) \;\; a(\hbox{\bf R})  =  0 \; , \;\; \kappa(\hbox{\bf R}) = 0
\Rightarrow  \Delta(F) = 0 , \\ \\
2.) \;\; a(\hbox{\bf R})  =  a_{0} (\hbox{\bf R} - c),
\; \kappa(\hbox{\bf R}) = \kappa_{0}(\hbox{\bf R} - c), \;
\gamma = \frac{1}{c + c^{-1}} + (b-\lambda) \Rightarrow  \\ \\
 (\hbox{\bf R} -c) \tilde{a}(\hbox{\bf R})  =
\frac{(\lambda - b)}{c} (\hbox{\bf R} -c), \;\;
 (\hbox{\bf R} -c) \tilde{\tilde{a}}(\hbox{\bf R}) =
\frac{b(\lambda - b)}{c^{2}} (\hbox{\bf R} -c) \equiv 0.
\end{array}
\end{equation}
Here $a_{0} \neq 0, \; \kappa_{0} \neq 0$ are constants.

Now, we deduce the covariant commutation relations
for the generators $F^{i}_{j}$
postulating the following natural quantum hyperplane condition
\begin{equation}
\lb{2.31}
(\hbox{\bf R} -c)(\F e)(\F' e') = 0 \; .
\end{equation}
Using (\ref{2.30a}) one can
obtain from (\ref{2.31}) the following relations
\begin{equation}
\lb{2.32}
 (\hbox{\bf R} -c) \F \hbox{\bf R} \F (\hbox{\bf R} + c^{-1}) = 0 \; .
\end{equation}
The commutation relations for the curvature
2-form $F^{i}_{j}$ have to be independent of the class of the
comodule $\{ e^{i} \}$ and therefore of the choice of the
parameter $c = \pm q^{\pm 1}$. So, we deduce from Eqs.(\ref{2.32})
the commutation relations
\begin{equation}
\lb{2.33}
 \hbox{\bf R}\F\hbox{\bf R}\F = \F\hbox{\bf R}\F\hbox{\bf R}  \; .
\end{equation}
These relations are known, first, as reflection equations \cite{K},
second, as the commutation relations for invariant vector fields
on $GL_{q}(N)$ \cite{Z,IPy} and, third, as the defining relations
for the braided algebras \cite{M}.

To complete the definition of the algebra $\O_{\bar{Z}}$
one can deduce
the following cross-commutation relation for $F$ and $A$:
\begin{equation}
\lb{2.34}
\F\hbox{\bf R}\A\hbox{\bf R} = \hbox{\bf R}\A\hbox{\bf R}\F \; .
\end{equation}
This is the simplest relation covariant under the coactions
presented in (\ref{2.20}) and (\ref{2.21a}) and allowing one to push
the operators $F$ through the operators $A$.

Thus, leaving aside the commutation relations with
generators $\{ e, \; de \}$, we come to the following
algebra with generators $A$ (1-form connection) and
$F = dA - A^{2}$ (2-form curvature):
\begin{equation}
\lb{2.35}
\begin{array}{rl}
\F\hbox{\bf R}\A\hbox{\bf R}  = \hbox{\bf R}\A\hbox{\bf R}\F \;  ,
& \hbox{\bf R}\F\hbox{\bf R}\F  = \F\hbox{\bf R}\F\hbox{\bf R}  \; , \\
\hbox{\bf R} \A \hbox{\bf R} \A + \A \hbox{\bf R} \A \hbox{\bf R}^{-1}
& = a(\hbox{\bf R}) (\F\hbox{\bf R} + \hbox{\bf R}^{-1}\F)
+ \kappa(\hbox{\bf R}) F^{0}\; ,
\end{array}
\end{equation}
where $a(\hbox{\bf R}) =  (\hbox{\bf R} - c) a_{0}$
and $\kappa(\hbox{\bf R}) =(\hbox{\bf R} - c) \kappa_{0}$
(see Eqs.(\ref{2.30c})).
Note, that for the case $a_{0} \neq 0$ and $\kappa_{0} \neq 0$
the consistence conditions
for the whole covariant algebra $\O_{\bar{Z}}$ give some additional
constraints on the generators of this algebra.
In particular, one can deduce
\begin{equation}
\lb{2.36}
(\hbox{\bf R} - c) \tilde{\F} \hbox{\bf R} e = 0,
\end{equation}
where $\tilde{F} = F - \frac{\kappa_{0}}{a_{0}(c+ c^{-1})} F^{0}$.

\section{$GL_{q}(N+1) / (GL_{q}(N) \otimes GL(1))$ noncommutative  geometry.}
\setcounter{equation}0

In this Section we present an explicit realization of such
covariant algebra $\O_{\bar{Z}}$ where parameters $a_{0}$,
$\kappa_{0}$ and additional relations (of the type (\ref{2.36}))
on the generators will be fixed. We consider the differential geometry
on the group $GL_{q}(N+1)$ \cite{Sch,Man1,Z,IPy} and interpret it as the
noncommutative geometry on the
total space of the principal fibre bundle
with the base space $GL_{q}(N+1)/(GL_{q}(N) \otimes GL(1)$ and
the structure group being $GL_{q}(N) \otimes GL(1)$.

Let us introduce $Z_{2}$-graded extension
of the $GL_{q}(N+1)$ quantum group (exterior Hopf algebra) with the
generators $\{ T^{I}_{J}, \; dT^{I}_{J} \} \;\; (I,J = 0,1, \dots N)$
satisfying the commutation relations (\ref{2.10})-(\ref{2.12}) where
$GL_{q}(N+1)$ $R$-matrix acts in the space $Mat(N+1) \times Mat(N+1)$.
Then, we consider the following left coaction of the group
$GL_{q}(N) \otimes GL(1)$ on the group $GL_{q}(N+1)$:
\begin{equation}
\lb{3.5}
T^{I}_{J} \rightarrow
 \left(
\begin{tabular}{c|c}
$ t $        &     $ 0 $   \\  \hline \\
$ 0 $        &     $ T^{i}_{k}    $    \\
\end{tabular}
\right) \otimes
 \left(
\begin{tabular}{c|c}
$T^{0}_{0}  $        &     $ T^{0}_{j} $ \\  \hline \\
$T_{0}^{k} $ &     $ T^{k}_{j}    $    \\
\end{tabular}
\right)
\end{equation}
where as usual $i,j,k = 1,2, \dots N$
and $t$ ($[t, \; T^{i}_{j} ] =0$) is a dilaton generator of
$GL(1)$. It is evident (from the commutation
relations for the $GL_{q}(N+1)$-generators) that the
elements $T^{i}_{j}$ generate
the quantum group $GL_{q}(N)$. The noncommutative coordinates
for the "base space" $GL_{q}(N+1) / (GL_{q}(N) \otimes GL(1))$
could be related with the generators $T^{0}_{i}$ and $T^{j}_{0}$.
For the Cartan 1-forms
on the $GL_{q}(N+1)$-group:
\begin{equation}
\lb{3.3}
\Omega^{I}_{J} = dT^{I}_{K}(T^{-1})^{K}_{J} = \left(
\begin{tabular}{c|c}
$\omega $        &     $ \Omega^{0}_{j} = < \bar{e} |_{j}$ \\  \hline \\
$\Omega_{0}^{i} = | e >^{i}$ &     $ A^{i}_{j}    $    \\
\end{tabular}
\right)
\end{equation}
the coaction (\ref{3.5}) is represented in
the form:
\begin{equation}
\lb{3.6}
\left(
\begin{tabular}{c|c}
$\omega $    &   $ < \bar{e} |$ \\  \hline \\
$ | e > $    &   $  A  $    \\
\end{tabular}
\right) \rightarrow
\left(
\begin{tabular}{c|c}
$\omega + dtt^{-1}$        &     $  < \bar{e} |T^{-1} t$ \\  \hline \\
$t^{-1} T | e > $ &     $T AT^{-1} + dTT^{-1}    $    \\
\end{tabular}
\right)
\end{equation}
where the short notation have been used (see e.g. (\ref{2.20})).
Comparing these transformations
with the transformations (\ref{2.17a}) and (\ref{2.20})
it becomes clear that the
Cartan 1-forms $|e>$ and
$A, \; \omega$ can be interpreted
as veilbein 1-forms and connection 1-forms respectively.
Then, the generators $< \bar{e} |$ are nothing but contragradient
veilbein 1-forms.
The Maurer-Cartan equation $d\Omega^{I}_{J} = \Omega^{I}_{K}
\Omega^{K}_{J}$
leads to the following constraints on the
noncommutative differential 1-forms $\O_{J}^{I}$:
\begin{equation}
\lb{3.7}
\left(
\begin{tabular}{c|c}
$d \omega - \omega^{2} -<\bar{e}|e> $
& $d < \bar{e} | - <\bar{e}|A - \omega <\bar{e}| $ \\  \hline \\
$d | e > - A | e > - | e >\omega  $
& $ dA -  A^{2} -  |e><\bar{e}|     $    \\
\end{tabular}
\right) = 0 \; .
\end{equation}
The $q$-deformed commutators for the noncommutative Cartan
1-forms (\ref{3.3}) is deduced from
the $N+1$-dimensional analog
of the relations presented in (\ref{2.23}).
Taking into account
Maurer-Cartan equations (\ref{3.7})
and using the notation (\ref{3.3}) we
rewrite these relations in the form:
\begin{equation}
\lb{3.8}
\hbox{\bf R} \A \hbox{\bf R} \A + \A \hbox{\bf R} \A \hbox{\bf R}^{-1}
= -\lambda (\hbox{\bf R} \F + \F \hbox{\bf R}^{-1} )
\end{equation}
\begin{equation}
\lb{3.9}
-e \A '  = \hbox{\bf R} \A \hbox{\bf R} e
+ \lambda \hbox{\bf R} (de - \A e) \; , \;\;
-\A ' \bar{e} = \bar{e} \hbox{\bf R} \A \hbox{\bf R}
+ \lambda (d\bar{e} -\bar{e}\A )\hbox{\bf R}
\end{equation}
\begin{equation}
\lb{3.10}
\bar{e} \hbox{\bf R} e  = -q e' \bar{e}', \;\;
\hbox{\bf R} e e'  = -q^{-1} e e' ,  \;\;
\bar{e}'\bar{e} \hbox{\bf R}   = -q^{-1} \bar{e}'\bar{e} ,
\end{equation}
\begin{equation}
\lb{3.10a}
 \omega^{2}=0, \;\; [\omega , e ]_{+} = [\omega , \bar{e} ]_{+} = 0, \;\;
[ A, \omega ]_{+} = q \lambda |e><\bar{e}| = q \lambda F.
\end{equation}
 Here we have also introduced the notation for the curvature 2-form:
\begin{equation}
\lb{3.12}
F = dA - A^{2} = |e><\bar{e}| =
-q^{-1} < \bar{e}|_{1} \hbox{\bf R} |e>_{1} \; .
\end{equation}
The last two equalities follow from Eqs.(\ref{3.7}) and (\ref{3.10})
and reveals the dependence of the curvature 2-forms and
the veilbein 1-forms.
Note, that for the such form (\ref{3.12}) of the curvature
one can directly prove the identity
(\ref{2.36}) (for $\kappa_{0} =0$) using the relations (\ref{3.10}).
Now, we find
(applying the commutation relations
(\ref{3.8})-(\ref{3.10a}) and Eq.(\ref{3.12})),
that the following relations for $F$ and $A$ are hold:
\begin{equation}
\hbox{\bf R} \F \hbox{\bf R} \F
=  \F \hbox{\bf R} \F \hbox{\bf R} \; , \;\;
\hbox{\bf R} \A \hbox{\bf R} \F = \F \hbox{\bf R} \A \hbox{\bf R}
+ \lambda (\hbox{\bf R} \F \omega - \F \omega \hbox{\bf R})
\lb{3.13}
\end{equation}
We would like to compare these relations with the relations
(\ref{2.35}) but on this stage we can not do it in view of the
appearing in (\ref{3.13}) the additional scalar generator
$\omega$ which is nothing but $GL(1)$-connection 1-form
(see (\ref{3.6})).
To exclude from the considerations these scalar connection 1-form
we introduce new total $GL_{q}(N) \otimes GL(1)$ connection:
\begin{equation}
\lb{3.15}
A_{t} = A - \omega I \; ,
\end{equation}
for which we have
\begin{equation}
\lb{3.15a}
\nabla_{t} e = \nabla_{t} \bar{e} = 0 \; ,
\end{equation}
(see (\ref{3.7})) and the corresponding curvature 2-form:
\begin{equation}
\lb{3.16}
F_{t} = q^{2} F - <\bar{e}|e> = q^{2}F + q^{1-N} \cdot F^{0}
\end{equation}
satisfy the conditions
\begin{equation}
\lb{3.16a}
F_{t}| e > = < \bar{e} | F_{t} = 0 \; .
\end{equation}
The scalar 2-form $F^{0} = Tr_{q}(F)$
in (\ref{3.16}) is defined by Eq.(\ref{2.20a})
and is invariant under the adjoint coaction
(\ref{2.21a}).
Finally, we find
from Eqs.(\ref{3.9})-(\ref{3.10a}) and (\ref{3.13})
that the elements
$\{ e, \; A_{t}, \; F \}$ generate
the following closed algebra:
\begin{equation}
\lb{3.17}
\begin{array}{c}
\hbox{\bf R} \F \hbox{\bf R} \F  = \F \hbox{\bf R} \F \hbox{\bf R} , \;\;
\hbox{\bf R} \A_{t} \hbox{\bf R} \F
= \F \hbox{\bf R} \A_{t} \hbox{\bf R} , \\
\hbox{\bf R} \A_{t} \hbox{\bf R} \A_{t}
+ \A_{t} \hbox{\bf R} \A_{t} \hbox{\bf R}^{-1}  =
a_{0} (\F \hbox{\bf R}^{-1} + \hbox{\bf R} \F)(\hbox{\bf R} -c), \\
-e \A_{t}' = \hbox{\bf R} \A_{t} \hbox{\bf R} e , \;\;
e \F ' = \hbox{\bf R} \F \hbox{\bf R} e
\end{array}
\end{equation}
where $a_{0} = 1-q^{2}$ and $c=-q^{-1}$.

Comparing the commutation relations (\ref{3.10})
and (\ref{3.17}) with the relations (\ref{2.1}), (\ref{2.26})
and (\ref{2.35}) one can infer that we have explicitly realized
the defining relations for
the covariant quantum algebra $\O_{\bar{Z}}$ of the type A.)
(\ref{2.26a}), (\ref{2.27a}) in terms of the algebraic objects
related to the $GL_{q}(N+1)/GL_{q}(N)\otimes GL(1)$-geometry.
To be precise we have to consider the algebra of the type (\ref{3.17})
with substitution $ F \leftrightarrow F_{t}$. The corresponding
defining relations are
\begin{equation}
\lb{3.18}
\begin{array}{c}
\hbox{\bf R} \F_{t} \hbox{\bf R} \F_{t}
= \F_{t} \hbox{\bf R} \F_{t} \hbox{\bf R} , \;\;
\hbox{\bf R} \A_{t} \hbox{\bf R} \F_{t}
= \F_{t} \hbox{\bf R} \A_{t} \hbox{\bf R} , \\
\hbox{\bf R} \A_{t} \hbox{\bf R} \A_{t}
+ \A_{t} \hbox{\bf R} \A_{t} \hbox{\bf R}^{-1}  =
a_{0} (\F_{t} \hbox{\bf R}^{-1} + \hbox{\bf R} \F_{t})(\hbox{\bf R} -c) +
\frac{ a_{0} (c + c^{-1}) (\hbox{\bf R} - c)}{q^{3} [N]_{q}} F_{t}^{0}, \\
-e \A_{t}' = \hbox{\bf R} \A_{t} \hbox{\bf R} e , \;\;
e \F_{t} ' = q^{-2} \hbox{\bf R}^{-1} \F_{t} \hbox{\bf R} e
\end{array}
\end{equation}
One can note that such kind algebras in view of the last relation of
(\ref{3.18}) were not presented in general consideration of
Sect.2. The explanation of this fact is that in the Sect.2
we essentially use the conditions $\nabla_{t} e \neq 0 \; ,
F_{t}| e > \neq 0$ which are not fulfilled here (see
(\ref{3.15a}) and (\ref{3.16a})). That is why we have not
received in the Sect.2 the
cross-commutation relations for $F$ and $e$
presented in ({\ref{3.18}).

\section{$GL_{q}(N)$-local co-invariants and Chern characters.}
\setcounter{equation}0

Our final aim is to define
composite elements ${\cal L}$ for the
extended algebra $\Omega_{\bar{Z}}$
which are co-invariant ${\cal L} \rightarrow 1 \otimes {\cal L}$
under the $GL_{q}(N)$ local transformations (\ref{2.4a}), (\ref{2.4b}),
(\ref{2.20}) and (\ref{2.21a}). We would like to interpret these elements
${\cal L}$ as noncommutative Lagrangians. However, we stress
that this interpretation is rather formal because the elements
${\cal L}$ are not the usual Lagrangians for certain field theories.
To write down such noncommutative Lagrangians
we further extend
the algebra $\Omega_{\bar{Z}}$ described in the Sect.2
by virtue of introducing
$Z_{2}$-graded contragradient comodule
$\left( \bar{e}_{i}, d\bar{e}_{j} \right)$
with the following commutation relations:
\begin{equation}
\lb{4.1}
\begin{array}{l}
\bar{e}' \cdot \bar{e}\hbox{\bf R} = c \bar{e}' \bar{e} \; , \;\;
(d \bar{e})' \bar{e} =
(\pm)c \bar{e}' (d \bar{e})\hbox{\bf R} , \cr\cr
(d \bar{e})' (d\bar{e})\hbox{\bf R} = -
{1 \over c} (d\bar{e})' (d\bar{e}) .
\end{array}
\end{equation}
Note that contragradient $q$-vectors have naturally appeared
in the context of the explicit example of the $GL_{q}(N)$-covariant
noncommutative geometry considered in the Sect.3.
The quantum group local (structure) transformation of the vector
($\bar{e}_{i}, d\bar{e}_{j}$) is expressed as the
following homomorphism
of the algebra (\ref{4.1}):
\begin{equation}
\lb{4.2}
\begin{array}{c}
( \bar{e}, \; d\bar{e} )
{\stackrel{g_{l}}{\longrightarrow } }
\left( (T^{-1})^{k}_{i} \otimes \bar{e}_{k} \; , \;\;
d(T^{-1})^{k}_{j} \otimes \bar{e}_{k} + (T^{-1})^{k}_{j}
\otimes d\bar{e}_{k}
 \right)
\equiv     \\ \\
\equiv (\bar{e},\; d\bar{e})
\cdot \pmatrix{T^{-1}, & -T^{-1}dT T^{-1} \cr 0 , & T^{-1} },
\end{array}
\end{equation}
where in the last equality of (\ref{4.2}) we have used the short
notation (see (\ref{2.20}), (\ref{2.21a}))
and the operators
$T^{i}_{j}$ and
$dT^{k}_{l}$ are the same as in Eqs.(\ref{2.10})-(\ref{2.12}).
The commutation relations for the coordinates
of the contragradient $q$-vectors
$\{ \bar{e}_{i}, \; d\bar{e}_{j} \}$
with the former generators of $\Omega_{\bar{Z}}$
can be found using
covariance of these relations under the gauge co-actions
(\ref{2.4a}), (\ref{2.4b}), (\ref{2.20}), (\ref{2.21a})
and (\ref{4.2}).
For example, one can assume the relations of the type
appeared in the explicit construction of the Sect.3:
\begin{equation}
\lb{4.3}
 e' \bar{e}' = c \bar{e} \hbox{\bf R} e \; , \;\;
(\pm) (de)' \bar{e}' = c (\bar{e} \hbox{\bf R} (de) +
\lambda \bar{e} \hbox{\bf R} \A \hbox{\bf R} e) \; ,
\end{equation}
\begin{equation}
\lb{4.4}
A' \bar{e} = (\pm) \bar{e} \hbox{\bf R} \A \hbox{\bf R} \; , \;\;
F' \bar{e} =  \bar{e} \hbox{\bf R} \F \hbox{\bf R} \; .
\end{equation}
These relations are not unique covariant relations for
the generators $\{ e, \bar{e},  A,  F, \dots \}$.
There are another choices corresponding to the another
noncommutative geometry. For example in our paper \cite{IP1}
we have proposed the noncommutative geometry with different
relations (\ref{4.3}).

Now one can define the co-invariant
elements of $\Omega_{\bar{Z}}$ transformed under the local
co-transformations as ${\cal L} \rightarrow 1 \otimes {\cal L}$.
For example, using
the noncommutative generators
$e^{i}, \; \bar{e}_{i}$ and $A_{i}^{j}$ we construct the
co-invariant
\begin{equation}
\lb{4.5}
{\cal L} = \bar{e}_{i} \left( de^{i} - A^{i}_{j}e^{j}\right) .
\end{equation}
We call these composite elements of the algebra $\Omega_{\bar{Z}}$
the noncommutative (algebraical) Lagrangians
bearing in mind the formal similarity of (\ref{4.5})
to the Lagrangians for the one dimensional
discrete gauge models (see e.g. \cite{FIG}).

In order to write down other local quantum group
co-invariants, it is convenient to use
the curvature 2-form $F$ transformed  as
the adjoint comodule (\ref{2.21a}).
As an example we present the noncommutative analogs of Chern
characters.
For this, let us consider the special
case of the closed algebra (\ref{2.35}) with the generators $A$ and $F$
where the parameters $a(\hbox{\bf R}) = 0$
and $\kappa(\R) = 0$. Here, as we have explained above,
$A^{i}_{j}$ are noncommutative analogs of connection 1-forms,
while $F^{i}_{j}$ are interpreted as curvature 2-forms.
In analogy with the classical case (see e.g. \cite{MiSt}),
we consider as invariant characters the following expressions:
\begin{equation}
\lb{4.1?}
C_{k} = Tr_{q}(F^{k}) = D^{i}_{j}F^{j}_{j_{1}} \cdots F^{j_{k-1}}_{i},
\end{equation}
where we have used the $q$-deformed trace
defined in (\ref{2.20a}).
Using (\ref{inv}) we immediately obtain that $2k$-forms
$C_{k}$ (\ref{4.1?}) are invariant under the adjoint coaction (\ref{2.21a}).
Moreover, $C_{k}$ are the closed $2k$-forms.
Indeed, from the Bianchi identities $dF=[A,F]$ we deduce
\begin{equation}
\lb{4.5?}
dC_{k} = Tr_{q}(AF^{k} - F^{k}A) = 0,
\end{equation}
where we have taken into account (see Eqs.(\ref{2.35}),
(\ref{4.3?}) and (\ref{2.a20}))
$$
\begin{array}{c}
Tr_{q}(AF^{k}) =
q^{-N}Tr_{q1}(Tr_{q2}(\R^{-1}\R \A \R \F^{k})) =
\\   \\
q^{-N}Tr_{q1}(Tr_{q2}(\F^{k}\R \A )) =
Tr_{q}(F^{k}A).
\end{array}
$$
We believe that $C_{k}$ have to be presented as the exact form
$C_{k}=dL^{(k)}_{CS}$, where the Chern-Simons $(2k-1)$-forms
$L^{(k)}_{CS}$ are represented as
\begin{equation}
\lb{4.6?}
L^{(k)}_{CS}=Tr_{q} \{ A(dA)^{k-1} + \frac{1}{h^{(k)}_{1}} A^{3}(dA)^{k-2} +
\dots + \frac{1}{h^{(k)}_{\dots}} A^{2k-1} \}
\end{equation}
and the constants $h^{(k)}_{i}$ depend on
the deformation parameter $q$.
We do not have explicit formulas for all parameters $h^{(k)}_{\dots}$
(in the classical case $q=1$ these formulas are known \cite{BZ}),
but for the case $k=2$ one can obtain a noncommutative analog of the
three-dimensional Chern-Simons term in the form:
\begin{equation}
\lb{4.7?}
L_{CS}^{(2)} = Tr_{q} \{ AdA - \frac{1}{h^{(2)}_{1}} A^{3} \} \; , \;\;
h^{(2)}_{1} = 1 + \frac{1}{q^{2} + q^{-2}} \; .
\end{equation}
We would like to note that it is
extremely interesting to write the Chern characters for the
general case of the algebra (\ref{2.35})
when the parameters $a(\R) \neq 0$ and $\kappa(\R) \neq 0$.

At the end of this Section we propose the way how to find
the algebraical Lagrangian corresponding to the
field theoretical Lagrangian
for the Einstein gravity. First, we take the four generators
of the underlying Zamolodchikov algebra
(\ref{2.1}) in the form of $2 \times 2$ matrix
$e^{ij} \; (i,j=1,2; \;\; e^{\dagger} = e)$
interpreted  as the spinorial representation
for the 4-dimensional veilbein 1-forms. The differential
complex $\Omega_{Z}$ for this algebra is the anticommuting
version $((\pm) = +1)$
of the differential complex for the $q$-Minkowski space \cite{Og,Kuli}
\begin{equation}
\R e\R e + e\R e\R^{-1} = 0\,, \label{4.8}
\end{equation}
\begin{equation}
\R\,de\,\R\,e - (\pm) e\,\R\,de\,\R = 0 \,, \label{4.9}
\end{equation}
\begin{equation}
\R\,de\,\R\,de - de\,\R\,de\,\R = 0 \,.  \label{4.10}
\end{equation}
Note that there is another consistent differential complex
with the choice of eq.(\ref{4.9}) in the form
$\R\,e\,\R\,de = (\pm) de\,\R\,e\,\R$.
Here we do not consider this possibility
which is absolutely parallel.
The factor $(\pm) = -1$ corresponds to the fermionic version of
the $q$-Minkowski space.
The algebra (\ref{4.8})-(\ref{4.10})
is covariant under the q-Lorentz global transformations
\begin{equation}
\lb{4.11a}
e \rightarrow T e \tilde{T}^{-1} \; ,
\end{equation}
\begin{equation}
\lb{4.11b}
de \rightarrow T de \tilde{T}^{-1} \; .
\end{equation}
where $\{ e, \; de \}$ commute with $\{ T, \; \tilde{T} \}$ and
elements of matrices $T$ and $\tilde{T} = (T^{\dagger})^{-1}$
are the generators
of the two $SL_{q}(2)$-groups with the following crossing-commutation
relations
\begin{equation}
\R T\tilde{T}' = \tilde{T}T'\R \; ,
\label{4.12}
\end{equation}
This formulation of the q-Lorentz group
have been proposed and investigated
in \cite{Car}-\cite{Kuli}. Using the $q$-trace (\ref{2.20a}) one
can construct from the generators $e^{ij}$ the contragradient
veilbein 1-forms $\bar{e}_{ij}$:
\begin{equation}
\label{4.13}
\bar{e}_{ij} = e^{ij} - q^{-1} Tr_{q}(e) \delta_{ij} \; .
\end{equation}
The co-transformation (\ref{4.11a}),(\ref{4.11b}) for $\bar{e}$ reads
\begin{equation}
\label{4.14}
\bar{e} \rightarrow \tilde{T} \bar{e} T^{-1} \; ,\;\;
d\bar{e} \rightarrow \tilde{T} d\bar{e} T^{-1} \; .
\end{equation}
Further we need the differential calculus on $SL_{q}(2)$.
Up to now we do not have the appropriate calculus on $SL_{q}(N)$
(see however \cite{FP}). Therefore we will consider the case of
extended Lorentz symmetry generated by $\Omega_{GL_{q}(2)}$.
In this case one can consider the local version of the transformation
(\ref{4.11b})
\begin{equation}
\label{4.16}
de \rightarrow dT e \tilde{T}^{-1} + T de \tilde{T}^{-1} - (\pm)
T e d\tilde{T}^{-1} \; ,
\end{equation}
where $\{ T, \; dT \}$ and $\{ \tilde{T}, \; d\tilde{T} \}$ are
two isomorphic $GL_{q}(2)$-exterior algebras (\ref{2.10})-(\ref{2.12})
with the cross-commutation relations defined by eq.(\ref{4.12})
and
\begin{equation}
\lb{4.17}
\begin{array}{rl}
 \R T\,d\tilde{T}' & = d\tilde{T}\,T'\R   \; , \\
 \R dT\,\tilde{T}' & = \tilde{T}\,dT'\R   \; ,  \\
\R dT\, d\tilde{T}' & = -  d\tilde{T}\, dT' \R\; ,
\end{array}
\end{equation}
Note that the formulas (\ref{2.10})-(\ref{2.12}),(\ref{4.12})
and (\ref{4.17}) for the $GL_{q}(N)$ $R$-matrix define the differential
complex on $GL_{q}(N,C)$.
Then one can introduce the covariant derivative
\begin{equation}
\lb{4.18}
(\nabla e) = de - Ae -e\tilde{A} \;
\end{equation}
where connection 1-forms $A$ and $\tilde{A}$ are transformed as
\begin{equation}
\lb{4.19}
A \rightarrow T A T^{-1} + dT T^{-1} \; , \;\;
\tilde{A} \rightarrow \tilde{T} \tilde{A} \tilde{T}^{-1} +
d\tilde{T} \tilde{T}^{-1} \; .
\end{equation}
For the consistence we demand that $\tilde{A} = A^{\dagger}$.
The corresponding curvature 2-forms
$F$ and $\tilde{F}$
are defined as usual
\begin{equation}
\lb{4.20}
F =dA -A^{2} \; , \;\;
\tilde{F} =d\tilde{A} -\tilde{A}^{2}\; .
\end{equation}
We assume that 2-forms $F$ and $\tilde{F}$
admit the expansion over the basis of the veilbein 1-forms
(cf. with (\ref{3.12}))
\begin{equation}
\lb{4.21}
F_{1} = Tr_{q2} (\bar{e}_{2} F_{12} e_{2}) \rightarrow
\tilde{F}_{1} = Tr_{q2} (e_{2} \tilde{F}_{12} \bar{e}_{2})  \; .
\end{equation}
The noncommutative scalar curvature
could be obtained as a real combination of the
coefficients $F_{12}, \; \tilde{F}_{12}$:
\begin{equation}
\lb{4.22}
{\cal F} = Tr_{q1} Tr_{q2} (F_{12} + \tilde{F}_{12}) \; ,
\end{equation}
and the corresponding algebraical version of the Einstein Lagrangian
reads
$$
{\cal L} = \mu(e^{ij}) \cdot {\cal F}
$$
where the invariant 4-dimensional real measure $\mu$
can be chosen in the form:
$$
\mu = i \left(
Tr_{q} (e \bar{e} e \bar{e}) - Tr_{q} (\bar{e} e \bar{e} e)
\right)  \; .
$$
Here $\bar{e}_{i}$ are contragradient veilbein 1-forms transformed
as in (\ref{4.14}).

\section{Discussion and Conclusion}
\setcounter{equation}0

To conclude the paper we would like to make some remarks
and comments.

\begin{enumerate}
\item We note that there is the realization of the
differential complex (\ref{2.10})-(\ref{2.12})
with the usual differential $d= dz\partial_{z} + d\bar{z}
\partial_{\bar{z}}$ over the classical
2-dimensional space $\{ z, \; \bar{z} \}$.
Indeed, let us consider the algebra
\begin{equation}
\lb{5.1}
\begin{array}{rcl}
\R T T'  & = & T T' \R \; , \\
T M' = \R M \R T & , & \bar{M} T' =  T' \R^{-1} \bar{M} \R^{-1} \; , \\
\R M \R M = M \R M \R & , &
\R^{-1} \bar{M}' \R^{-1} \bar{M}'
= \bar{M}' \R^{-1} \bar{M}' \R^{-1} , \\
\; \left[ \bar{M}  , \; M' \right] & = & 0 \; ,
\end{array}
\end{equation}
where as usual $M = M_{1}$ and $M' = M_{2}$ etc. Then one
can prove that the operators
\begin{equation}
\lb{5.2}
\begin{array}{c}
T(z, \; \bar{z} ) = exp(zM) T exp(\bar{z}\bar{M}) , \\
dT(z, \; \bar{z} ) = dz (\partial_{z} T)
+ d\bar{z} (\partial_{\bar{z}} T) = dz MT + d\bar{z} T \bar{M}
\end{array}
\end{equation}
satisfy the commutation relations (\ref{2.10})-(\ref{2.12}).
The generators $ \{ e^{i}, \; (de^{i}) \}$ of the exterior algebra
$\O_{Z}$ (\ref{2.1}) for $c= q$ can be realized now as
columns of the quantum matrices $T^{i}_{j}(z,\bar{z})$ and
$dT^{i}_{j}(z, \bar{z})$. In this sense we indeed can consider
Eqs.(\ref{2.4a}),(\ref{2.4b}) as a local co-transformations
where $\{ z, \; \bar{z} \}$ are coordinates of the space-time.
We stress also that Eqs.(\ref{5.1}) and (\ref{5.2}) remind the
formulas appeared in the framework of the Hamiltonian quantizing
of the WZWN models (see e.g. \cite{Fadd} and references therein)
and related toy model \cite{AlFad}.

\item
Another attractive possibility is the choice of the noncommutative
space-time
isomorphic to the space of the quantum group e.g. $GL_{q}(N)$. In this
case it is tempting to explore monopole-instanton type gauge potential
1-forms
\begin{equation}
\lb{5.2'}
A^{i}_{j}=dT^{i}_{k} M^{k}_{l}(Z)(T^{-1})^{l}_{j} =
dT^{i}_{k}(T^{-1})^{l}_{j} M^{k}_{l}(Z) \; ,
\end{equation}
where $Z=det_{q}T$ and
$([M(Z) ,\; T ] = 0, \;
M(Z)\, dT  = dT\, M(q^{2}Z))$.
Substituting (\ref{5.2'}) in the anticommutation relations (\ref{2.23})
we obtain that $M$ satisfy reflection equation:
$$
M(q^{2}Z)\R^{-1} M(Z) \R^{-1} -
\R^{-1}M(q^{2}Z)\R^{-1}M(Z) = 0 \; .
$$

\item For arbitrary invertible Yang-Baxter $R$-matrix satisfying
the characteristic equation (generalization of (\ref{2.2}))
\begin{equation}
\lb{5.3}
(\R - \lambda_{1})(\R - \lambda_{2}) \cdots
(\R - \lambda_{m}) = 0, \;\; (\lambda_{i} \neq \lambda_{j} \;\; if
\;\; i \neq j )
\end{equation}
one can introduce \cite{Hlav} the set of the quantum hyperplanes and
covariant differential calculi on them.
Namely, for each eigenvalue $\lambda_{k}$
we define the exterior algebra
$\{ e, \; (de) \}$ with the commutation relations
\cite{Hlav}
\begin{equation}
\lb{5.4}
\begin{array}{rcl}
\prod\limits_{j \neq k}
\frac{\hbox{({\bf R} - $\lambda_{j}$)}}{\hbox{$(\lambda_{k}
-\lambda_{j})$}} e e' & \equiv &
\hbox{\bf P}_{k} e e'  = 0 \\
\R (de) e' & = & -\lambda_{k} e (de)' \; , \\
\R (de) (de)' & = & \lambda_{k} (de) (de)' ,
\end{array}
\end{equation}
We choose two variants of the hyperplanes
related to the eigenvalues $\lambda_{k}$ and $\lambda_{i}$ for which
projectors $P_{k}$ and $P_{i}$ are $q$-analogs of a
symmetrizer and an antisymmetrizer (fermionic and bosonic hyperplanes).
Then we deduce the commutation relations for $T$ and $dT$
substituting the transformations (\ref{2.4a}), (\ref{2.4b})
into these two variants of the relations (\ref{5.4}).
Surprisingly these relations
coincide with the relations (\ref{2.10})-(\ref{2.12})
for $\lambda_{k}\lambda_{i} = 1$ and,
as it can be easily shown, such differential
complex is not consistent for $m > 2$,
e.g. for the quantum groups such as $SO_{q}(N)$ and $SP_{q}(2N)$
for which $m=3$.
Our conjecture is that the consistent
differential complex
for quantum groups with general $R$-matrices satisfying (\ref{5.3})
can be represented in the form
(cf. with formulas presented in \cite{CaC})
\begin{equation}
\lb{5.5a}
\R T T'  =  T T' \R \; ,
\end{equation}
\begin{equation}
T (dT)'  =  \sum_{k,j=1}^{m} \alpha_{kj}
\hbox{\bf P}_{k} (dT) T' \hbox{\bf P}_{j} - (dT) T' \; ,
\lb{5.5b}
\end{equation}
\begin{equation}
\sum_{k,j=1}^{m} \alpha_{kj}
\hbox{\bf P}_{k} (dT) (dT)' \hbox{\bf P}_{j}  =  0 \; .
\lb{5.5c}
\end{equation}
Here the differential $d$ satisfies the undeformed
graded Leibnitz rule,
the coefficients $\alpha_{kj}=0,1 \;\; (k \neq j)$
and $\alpha_{k} \equiv \alpha_{kk}$ have to be fixed from the
diamond condition (the unique
lexicographic ordering of cubic monomials)
for the algebra (\ref{5.5a})-(\ref{5.5c}). In particular one can
deduce the following condition on $\alpha_{k}$
$$
[ X( \Omega ), \; \R ] = 0
$$
where
$X ( \Omega ) = (1 - \sum_{k} \alpha_{k} \hbox{\bf P}_{k}') \Omega_{1} +
\sum_{k,l} \alpha_{k}\alpha_{l}
\hbox{\bf P}_{k}'\hbox{\bf P}_{l} \Omega_{1}
\hbox{\bf P}_{l}\hbox{\bf P}_{k}'$.
Note, that the algebra (\ref{5.5a})-(\ref{5.5c}) is an exterior Hopf
algebra with the structure maps defined in (\ref{2.13}).

\item Now we make some notes about Brzezinski theorem \cite{B} and
its application to the construction of the
quantum group covariant noncommutative geometry.

Let $({\cal A}, \; \Delta, \; {\cal S}, \epsilon)$  be a Hopf algebra
and $(\Gamma, \; d)$ - first order differential calculus on ${\cal A}$,
where $\Gamma$ is a space of 1-forms on ${\cal A}$, while $d$ is a
differential mapping which is nilpotent $d^{2}=0$ and satisfies
graded Leibnitz rule. Denote the basic elements of ${\cal A}$
(including unity) as $\{ t_{i}, \; t_{0} = 1 \}$ and define
\begin{equation}
\lb{5.6a}
t_{i}t_{j} = m^{k}_{ij} t_{k} \; ,
\end{equation}
\begin{equation}
\lb{5.6b}
\Delta(t_{i})= \Delta^{kj}_{i} t_{k}\otimes t_{j} \; ,
\end{equation}
\begin{equation}
\lb{5.6c}
{\cal S}(t_{i})= S_{i}^{j} t_{j} \; .
\end{equation}
The comultiplication $\Delta$ is a homomorphic
mapping for the algebra (\ref{5.6a}) and therefore
we have the following condition on the structure constants
\begin{equation}
\lb{5.7}
\Delta^{kn}_{i} \Delta^{ql}_{j}
m^{p}_{kq} m^{r}_{nl} = m^{k}_{ij} \Delta^{pr}_{k} \; .
\end{equation}
Let us choose in $\Gamma$ the basis of independent
1-forms $\{ \omega_{\alpha} \}$
defined by the relations
\begin{equation}
\lb{5.8}
dt_{i} = (\chi^{\alpha})_{i}^{j} \omega_{\alpha} t_{j} \; .
\end{equation}
where $\chi^{\alpha}$ are some numerical matrices.
Each element in $\Gamma$ can be uniquely represented
in the form
$\sum a_{\alpha} \omega_{\alpha}$ or
$\sum \omega_{\alpha} b_{\alpha},\;\;
(a_{\alpha}, \; b_{\alpha} \in {\cal A})$ and therefore we have
to be able to commute $\{t_{m}\}$ and $\{ \omega_{\alpha} \}$:
\begin{equation}
\lb{5.9}
t_{n} \omega_{\beta} = (F^{\alpha}_{\beta})^{k}_{n} \omega_{\alpha}t_{k} \; ,
\end{equation}
where
$$
(F^{\alpha}_{\beta})^{k}_{n} = \eta_{\gamma\beta}
\left( (\chi^{\alpha})^{l}_{r} m^{r}_{nj}
(\chi^{\gamma})^{j}_{l} \delta^{k}_{0}
- Tr(\chi^{\gamma}) (\chi^{\alpha})^{k}_{n}  \right) ,
$$
$$
( \eta_{\alpha\beta}\eta^{\beta\gamma}=\delta^{\gamma}_{\alpha} \;\;
{\rm and} \;\; \eta^{\alpha\beta}=Tr(\chi^{\alpha}\chi^{\beta}) )
$$
are again some invertible numerical matrices.
The corresponding commutation relations for the basis of 1-forms
(in other words
the definition of the exterior product $\omega \wedge \omega$)
can be easily deduced by the differentiation of Eq.(\ref{5.9})
\begin{equation}
\lb{5.10}
\left[ \chi^{\alpha}
\omega_{\alpha} \; , \;\; F^{\gamma}_{\beta} \omega_{\gamma} \right]_{+} =
\left( F^{\alpha}_{\beta} f^{\gamma'\delta}_{\alpha} -
f^{\alpha\xi}_{\beta}F^{\gamma'}_{\alpha}
F^{\delta}_{\xi} \right) \omega_{\gamma'} \omega_{\delta} \; .
\end{equation}
One can guarantee that there are no other quadratic relations
on $\omega_{\alpha}$ since we choose these 1-forms
as independent. We imply in Eq.(\ref{5.10})
the exterior products of the differential forms
and introduce structure constants $f^{\alpha\beta}_{\gamma}$
appeared in the Maurer-Cartan equation
\begin{equation}
\lb{5.11}
d \omega_{\alpha} =
f^{\beta\gamma}_{\alpha} \omega_{\beta} \wedge \omega_{\gamma} \; .
\end{equation}
Comparing this relation with
the differential of Eq.(\ref{5.8}) one can
express $f^{\beta\gamma}_{\alpha}$ in terms of the matrices
$\chi^{\gamma}$.

The relations (\ref{5.6a}), (\ref{5.9}) and (\ref{5.10}) are defining
relations for the exterior algebra $\Omega = \bigoplus\limits_{n=0}
\Omega^{(n)}$ of ${\cal A}$. Here
$\Omega^{(0)}={\cal A}$, $\Omega^{(1)}=\Gamma$ and $\Omega^{(n)}$
denotes the space of n-forms. Now let us consider the mapping
$\Delta'$: $\Omega \rightarrow \Omega \otimes \Omega$
where $\otimes$ is a graded tensor product and
$\Delta'({\cal A}) \equiv \Delta ({\cal A})$.
Define the action of $d$ on $\Omega \otimes \Omega$
as
$$
d (\Omega^{(n)} \otimes \Omega^{(k)}) = d \Omega^{(n)} \otimes \Omega^{(k)}
+ (-1)^{n} \Omega^{(n)} \otimes d \Omega^{(k)}.
$$
Our proposition is that if the mapping $\Delta'$
(co-action) commutes with $d$:
\begin{equation}
\lb{5.12}
d \Delta' = \Delta' d
\end{equation}
and the relations (\ref{5.9}) are covariant  under the co-action
$\Delta'$, then the differential complex
(\ref{5.6a}), (\ref{5.9}) and (\ref{5.10}) defines the exterior Hopf
algebra of ${\cal A}$.

\underline{\bf Proof:} First, we note that from the condition
(\ref{5.12}) we obtain the explicit definition of $\Delta'$:
\begin{equation}
\lb{5.13}
\begin{array}{c}
\Delta'(t_{i}) = \Delta(t_{i}) \\
\Delta'(dt_{i}) = d \Delta'(t_{i}) =
\Delta^{kj}_{i}(dt_{k} \otimes t_{j} + t_{k} \otimes d t_{j}) \; .
\end{array}
\end{equation}
The co-action on the higher differential forms $\Omega^{(n)}$
can be derived from (\ref{5.13}).
 From the covariance of the relations (\ref{5.9})
it is not hard to show
(applying Leibnitz rule and the condition (\ref{5.12}))
that the relations (\ref{5.10}) are also covariant under the
co-action (\ref{5.13}). The coassociativity of $\Delta$
$$
\Delta^{kj}_{i} \Delta^{ln}_{j} = \Delta^{jn}_{i} \Delta^{kl}_{j}
$$
leads to the coassociativity
of $\Delta'$ (\ref{5.13}). Thus, $\Delta'$ is a coproduct for
${\cal A} \oplus \Gamma$ and therefore for $\Omega$.
Finally, we define the extended versions of the antipode ${\cal S}'$
and the counite $\epsilon'$ for the exterior algebra $\Omega$
by means of the relations
\begin{equation}
\lb{5.14}
\begin{array}{rl}
{\cal S}'(t_{i}) = {\cal S}(t_{i}) , &
{\cal S}'(dt_{i}) = d {\cal S}(t_{i}) , \; \\
\epsilon'(t_{i}) = \epsilon(t_{i}) , &
\epsilon'(dt_{i}) = d \epsilon(t_{i}) = 0.
\end{array}
\end{equation}
All axioms for ${\cal S}'$ and $\epsilon'$ follow from the corresponding
axioms for ${\cal S}$ and $\epsilon$.

This proposition immediately implies Brzezinski theorem \cite{B}
since the bicovariance for $(\Gamma, \; d)$ is nothing but
the covariance of the relations
(\ref{5.6a}), (\ref{5.9}) and (\ref{5.10}) with respect to the
left $\Phi_{L}$ and right $\Phi_{R}$ coactions on ${\cal A} \oplus \Gamma$
\begin{equation}
\lb{5.15}
\begin{array}{c}
\Phi_{L,R}(t_{i}) = \Delta(t_{i}) \\
\Phi_{L}(dt_{i}) = \Delta^{kj}_{i} t_{k} \otimes d t_{j} \; ,
\Phi_{R}(dt_{i}) = \Delta^{kj}_{i} dt_{k} \otimes  t_{j} \; ,
\end{array}
\end{equation}
and therefore relations (\ref{5.9}) are also covariant under the coaction
(\ref{5.13}).

Now we consider the left coaction of the exterior Hopf
algebra $\Omega$ on a left comodule
represented by some exterior algebra $\Omega_{Z}$:
\begin{equation}
\lb{5.16}
\begin{array}{c}
x_{\alpha} \rightarrow (C^{i})^{\beta}_{\alpha} t_{i} \otimes x_{\beta}, \\
dx_{\alpha} \rightarrow (C^{i})^{\beta}_{\alpha}
( dt_{i} \otimes x_{\beta} +
t_{i} \otimes dx_{\beta} ).
\end{array}
\end{equation}
Here $\{ x_{\alpha} , \; d x_{\alpha} \}$
are generators of $\Omega_{Z}$ and
matrices $C^{i}$ represent the dual object:
$(C^{i})^{\beta}_{\alpha} (C^{j})^{\gamma}_{\beta} =
\Delta^{ij}_{k} (C^{k})^{\gamma}_{\alpha}$. If we extend
the algebra $\Omega_{Z} \rightarrow \Omega_{\bar{Z}}$
by adding new generators
$A_{\alpha}^{\beta}$ such that
$A_{\alpha}^{\beta} \in \Omega^{(1)}_{\bar{Z}}$
and introduce a new differential $\nabla x_{\alpha} = d x_{\alpha} -
A_{\alpha}^{\beta} x_{\beta}$ transformed covariantly under
(\ref{5.16})
\begin{equation}
\lb{5.17}
\nabla x_{\alpha} \rightarrow
(C^{i})^{\beta}_{\alpha} t_{i} \otimes \nabla x_{\beta},
\end{equation}
then we interpret $A_{\alpha}^{\beta}$
as connection 1-forms. The definition of the curvature 2-forms
is evident. One can try to construct
the cross-product of the algebras $\Omega$ and $\Omega_{\bar{Z}}$
and obtain a new exterior Hopf algebra $G$ for which $\Omega$ will be
a Hopf subalgebra. In this case $A_{\alpha}^{\beta}$
could be realized as right-covariant 1-forms on $G$.
Just this realization have been done in the Sect.3
where $\Omega \equiv \Omega_{GL_{q}(N)}$ and
$G \equiv \Omega_{GL_{q}(N+1)}$.
So we see that in principal the algebraical constructions
of the Sections 2 and 3 could be adapt with the case of
the arbitrary exterior Hopf algebra.

\item Finally, we would like to stress that there are many
variants of the quantum group covariant commutation relations
for connections, curvatures, veilbeins etc. For each
variant (and for the same quantum group of covariance)
one can obtain different noncommutative geometries. Therefore
the structure co-group (the group of the covariance) is not
define the noncommutative geometry uniquely. Indeed, we can embed
the structure quantum group $\Omega$ in various large
algebras $G$ and
correspondingly to obtain various geometrical structures.
For example, one can consider the embedding of the
structure group $\Omega = \Omega_{GL_{q}(N)}$
in the arbitrary group $\Omega_{GL_{q}(M)}$ for
$M > N+1$. Obviously this will be the generalization of the
noncommutative geometry for $M=(N+1)$
presented in Sect.3.

It seems that all these ideas very closely related to the concept of the
noncommutative geometry on the quantum
principal fibre bundles \cite{BMa}. However, we stress
that we have not done the sequential analyses of these relations.
It would be very interesting to interpret the
quantum group covariant noncommutative
geometries as geometries on noncommutative
principal fibre bundles.

\end{enumerate}

\section*{Acknowledgments} The author would like to thank
I.Ya.Aref'eva, G.E.Arutyunov, L.Castellani,
P.P.Kulish, Sh.Majid, A.A.Vladimirov, A.T.Filippov
and especially Z.Popowicz and P.N.Pyatov for
the helpful discussions,
constructive criticism and interest to this work.

This work was supported in part by the
Russian Foundation of Fundamental Research (grant 93-02-3827).

%\newpage


\begin{thebibliography}{99}
\bibitem{Con} A.Connes, \it Geometrie non commutative,
\rm (Intereditions, Paris, 1990).
\bibitem{FRT}  L.D.Faddeev, N.Yu.Reshetikhin and L.A.Takhtadjan,
\it Algeb.Anal. \bf 1 \rm No.1 (1989)178.
\bibitem{D-M}  V.G.Drinfeld,
\it Quantum Groups, \rm in Proc. Inter. Congress of Mathematics
   vol.1(Berkley 1986)798;
   M.Jimbo, \it Lett.Math.Phys. \bf 10 \rm (1985) 63;
ibid. \bf 11 \rm (1986)247;
   S.L.Woronowicz, \it Comm.Math.Phys.\bf 111\rm(1987)613;
 Yu.Manin,
\it Quantum Groups and Noncommutative Geometry, \rm Montreal University
   Prep. CRM-1561 (1989); Comm.Math.Phys. \bf 122 \rm(1989)163.
\bibitem{W} S.L.Woronowicz,\it Comm.Math.Phys.\bf 122\rm(1989)125.
\bibitem{WJ} L.D.Faddeev, \it Lectures on Int. Workshop "Interplay
between Mathematics and Physics", \rm Vienna, 1992 (unpublished);
    B.Jur\v{c}o, \it Lett.Math.Phys. \bf 22 \rm(1991)177;
    T.Brzezinski and Sh.Majid,
\it Lett.Math.Phys. {\bf 26} \rm (1992)67;
 P.Aschieri and L.Castellani, \it Int.J.Mod.Phys. \bf A8 \rm (1993)1667;
F.M\"{u}ller-Hoissen, \it J.Phys. A: Math.Gen. {\bf 25} \rm (1992)1703.
\bibitem{CaC} U.Carow-Watamura , M.Schlieker, S.Watamura and W.Weich,
 \it Comm.Math. \\ Phys. \bf 142 \rm(1991)605; L.Castellani and
M.A.R.-Monteiro \it Phys.Lett. \bf B314 \rm(1993)25.
\bibitem{Z} B.Zumino, \it Introduction to the Differential
Geometry of Quantum Group, \rm Prep. LBL-31432 (1991);
    P.Schupp, P.Watts and B.Zumino, \it Lett.Math.Phys. \bf 25
		\rm(1992)139.
\bibitem{IPy}  A.P.Isaev and P.N.Pyatov,
\it Phys.Lett. \bf A179 \rm (1993)81.
\bibitem{IP1} A.P.Isaev and Z.Popowicz,
\it  Phys.Lett. \bf B307 \rm (1993)353.
\bibitem{GT}
I.Ya.Aref`eva and I.V.Volovich, \it Mod.Phys.Lett. \bf A6 \rm (1991) 893;
\it Phys.Lett. \bf B264 \rm(1991) 62;
 A.P.Isaev and Z.Popowicz, \it Phys.Lett. \bf B281 \rm(1992) 271;
  T.Brzezinski and Sh.Majid,
\it Quant. Group Gauge Theory on Classical
    Spaces, \rm Prep. DAMPT/92-51 (1992);
 D.Bernard, \it Suppl.Progr.Theor.Phys. \bf 102 \rm (1992) 49;
K.Wu and R.J.Zhang, \it  Comm.Theor.Phys.
\bf 17 \rm (1992)175;
V.Akulov, V.Gershun and A.Gumenchuk,
\it JETP Letters, \bf 56 \rm (1992)180.
\bibitem{BMa} T.Brzezinski and Sh.Majid,
\it Comm.Math.Phys. \bf 157 \rm (1993)591.
\bibitem{CW} L.Castellani, \it Phys.Lett., \bf B292 \rm (1992)93;
  S.Watamura,
 \it Quantum Deformations of BRST Algebra, \rm Heidelberg Univ.
   Prep. HD-THEP-92-39(1992);
 \it Bicovariant Differential Calculus and q-Deformation
 of Gauge Theory, \rm Heidelberg Univ.
   Prep. HD-THEP-92-45(1992).
\bibitem{AA} I.Ya.Aref'eva and G.E.Arutyunov,
\it Steklov Math.Inst. Prep., \rm SMI-4-93(1993).
\bibitem{Is} A.P.Isaev, \it "$GL_{q}(N)$-covariant
Noncommutative Geometry", \rm in Proceed. of the XXIIth Conference
"Differential Geometric Methods in Theoretical Physics" (Mexico,
20-25 September, 1993).
\bibitem{IPy2} A.P.Isaev and P.N.Pyatov, \it "Covariant
Differential Complexes on Quantum Linear Groups", \rm
Preprint JINR, Dubna, E2-93-416 (1993).
\bibitem{B} T.Brzezinski, \it Lett.Math.Phys. \bf 27\rm(1993)287.
\bibitem{IV} A.P.Isaev and A.A.Vladimirov, \it "$GL_{q}(N)$-Covariant
Braided Differential Bialgebras", \rm Dubna preprint JINR E2-94-32,
hep-th/9402024.
\bibitem{WZ} J.Wess and B.Zumino, \it Nucl.Phys.
(Proc. Suppl.) \bf B18 \rm(1990)302.
\bibitem{Sch} A.Sudbery, \it Phys.Lett. \bf B284 \rm (1992)61;
A.Sudbery, \it Math.Proc.Camb.Phyl.Soc. \bf 114 \rm (1993)111;
A.Schirrmacher, \it Remarks on use of R-matrices,
\rm in Proc. of the 1st Max Born Symposium, eds. R.Gielerak et al.
(Kluwer Acad. Publ., 1992) p.55.
\bibitem{Man1} Yu.I.Manin, \it Theor.Math.Fiz., \bf 92 \rm (1992)425.
\bibitem{IM} A.P.Isaev and R.P.Malik, \it Phys.Lett. \bf B280
\rm (1992)219.
\bibitem{K} P.Kulish and R.Sasaki, \it Progr.Theor.Phys.,
\bf 89 \rm(1993) 741.
\bibitem{M} Sh.Majid, \it J.Math.Phys. \bf 32 \rm (1991)3246;
\it ibid., \bf 34 \rm (1992)1176.
\bibitem{FIG} A.Filippov, D.Gangopadhyay and A.P.Isaev,
\it Int.J.Mod.Phys. \bf A7 \rm (1992)2487.
\bibitem{MiSt} J.W.Milnor and J.D.Stasheff, \it Characteristic Classes,
\rm Princeton Univ. Press and Univ. of Tokyo Press (1974).
\bibitem{BZ} B.Zumino, Wu Yong-Shi and A.Zee, \it Nucl.Phys.
\bf B239 \rm (1984)477; B.Zumino, \it Nucl.Phys. \bf 253B \rm(1985)477.
\bibitem{Car} U.Carow-Watamura , M.Schlieker, M.Scholl and
S.Watamura, \it Z.Phys.C \bf 48 \rm (1990)159.
\bibitem{Og} O.Ogievetsky, W.B.Schmidke, J.Wess and B.Zumino,
\it Comm.Math.Phys. \bf 150 \rm (1992)495.
\bibitem{Kuli} J.A. de Azcarraga, P.P.Kulish and F.Rodenas,
Valencia University preprint FTUV 93-36, 1993;
P.P.Kulish, Valencia University preprint FTUV 93-54, 1993.
\bibitem{FP} L.D.Faddeev and P.N.Pyatov, \it
"The Differential Calculus on Quantum Linear Group", \rm hep-th/9402070.
\bibitem{Fadd} L.D.Faddeev, \it in Filds and Particles, Proceedings
of the XXIX Winter School in Nuclear Physics, \rm eds. H.Mitter and
W.Schweiger, Schladmig, Austria (Springer, 1990);
A.Alekseev, L.D.Faddeev and M.Semenov-Tian-Shansky,
\it Comm.Math.Phys. \bf 149 \rm (1992)335;
A.P.Isaev, \it Theor.Mat.Phys., \bf 71 \rm No.3 (1987)616.
\bibitem{AlFad} A.Yu.Alekseev and L.D.Faddeev, \it Commun.Math.Phys.
\bf 141 \rm (1991)413.
\bibitem{Hlav} L.Hlavaty, \it J.Phys. A: Math.Gen. \bf 25 \rm(1992)485.

\end{thebibliography}
\end{document}